\documentstyle[11pt,fullpage,doublespace,epsf]{article}
 \DeclareMathSizes{11}{19}{13}{9}   

\makeatletter
\@addtoreset{equation}{section}
\makeatother

\begin{document}

\title{Cosmological Constant Seesaw\\ in 
\\Quantum Cosmology}
\author{Michael McGuigan\\Brookhaven National Laboratory\\Upton NY 11973\\mcguigan@bnl.gov}
\date{}
\maketitle

\begin{abstract}Recently a phenomenological relationship for the observed cosmological constant has been discussed by Motl and Carroll in the context of treating the cosmological constant as a $2\times 2$ matrix but no specific realization of the idea was provided. We realize a cosmological constant seesaw mechanism in the context of quantum cosmology. The main observation used is that a positive cosmological constant plays the role of a $Mass^2$ term in the Wheeler DeWitt (WDW) equation. Modifying the WDW equation to include a  coupling between two universes, one of which has planck scale vacuum energy and another which  has  vacuum energy at the supersymmetry breaking scale before mixing, we obtain the relation $\lambda = (10TeV)^8/M_{Pl}^4$ in a similar manner to the usual seesaw mechanism. We discuss how the picture fits in with our current understanding of string/M-theory cosmologies. In particular we discuss how these results might be extended in the context of exact wave functions of the universe derived from certain string models. \end{abstract}

\section{Introduction}

The small value of the cosmological constant in natural units $\lambda = 10^{-120} M_{Pl}$
is one one of the most perplexing problems in physics. The problem has proved so daunting that some have resorted to anthropic explanations involving $10^{500}$ universes. Recently a phenomenological relationship for the observed cosmological constant has been discussed by Motl and Carroll \cite{Motl}\cite{Carroll}. The basic idea was to treat the cosmological constant as a small eigenvalue of a $2\times 2$ matrix. However a universe has only one cosmological constant which is a number and not a matrix and it wasn't obvious how to implement the idea.  

One way might be to introduce mixing between two universes with different cosmological constants. A possible approach is to couple Wheeler-DeWitt (WDW) equations for  each universe together in order to effect topology change. 
Schematically we have:

\[
K_{12} (\Sigma ,\Sigma ') \sim \int\limits_\Sigma ^C {Dge^{iS_{\lambda _1 } } } \int\limits_C^{\Sigma '} {Dge^{iS_{\lambda _2 } } }  \sim \Phi _{1C} (\Sigma )\Phi _{2C}^* (\Sigma ')
\]
where $K_{12}$ is a transition amplitude between two
 spatial hypersurfaces $\Sigma$ and $\Sigma'$, $C$ is an interpolating hypersurface, $S_{\lambda_1}$ and $S_{\lambda_2}$ are actions with cosmological constant $\lambda_1$ and $\lambda_2$,  and $\Phi_1$ and $\Phi_2$ are solutions to the WDW equations $(\Delta  + \lambda_1 )\Phi_1  = 0$  and $(\Delta  + \lambda_2 )\Phi_2  = 0$ associated with these cosmological constants.

Also it  has been known for  some time that the cosmological constant plays the role of a $Mass^2$ in the WDW equation for a certain choice of variables \cite{McGuigan:1989yb}\cite{Russo:2004am}\cite{Townsend:2004zp}. Thus including mixing between universes might effectively turn a  $Mass^2$ term  into a $Mass^2$ matrix. If one solves the coupled set of WDW equations by diagonalizing this matrix,  the solution could describe a universe with a phenomenologically viable value of the cosmological constant obeying a seesaw relation $\lambda = (10TeV)^8/M_{Pl}^4$. The purpose of this paper is to investigate this scenario.

This paper is organized a follows. In section 2 we describe a basic description of the cosmological constant seesaw mechanism in the quantum cosmology of gravity coupled a to a system of scalar fields.
In section 3 we apply this approach to axion/dilaton gravity used in string theory. In section 4 discuss the implementation in M-theory inspired gravity models and contrast the cosmological seesaw approach with the Bousso-Polchinski mechanism. In section section 5 we discuss how to realize the cosmological seesaw in ultraviolet finite theories like $2+1$ dimensional gravity and certain string models within which one  can perform exact calculations.

\section{A Tale of Two Universes}

In this section we present the basic model leading to a cosmological constant seesaw mechanism. The main elements we shall need is a description of the Wheeler-DeWitt equation in quantum cosmology, the role of the cosmological constant as a $Mass^2$ term in the the WDW equation and some discussion of how topology change can effect these equations.

\subsection{Wheeler-DeWitt equation}

We first develop the equations for quantum cosmology for gravity coupled to a system of scalar fields. Other more specific choices of matter will be discussed in the following sections. We mainly follow the treatment of \cite{Fischler:1989ka}. The Einstein-Hilbert action coupled to  a system of scalar fields $\phi^I$ is given by

\begin{equation}
S = \frac{1}{2}\int {d^4 x(M_{Pl}^2 \sqrt { - g} } R - \sqrt { - g} \lambda  - \partial \phi ^I \partial \phi ^I )
\end{equation}
We choose the ansatz for the metric tensor given by:

\[
d\ell ^2  =  - N^2 (t)dt^2  + a^2 (t)d\Omega _3^2 
\]
Here $d\Omega _3^2$ is the line element for a space of constant spatial curvature.
Under the ansatz the action becomes

\begin{equation}
S = \frac{1}{2}\int {dta^3 N( - 9M_{Pl}^2 (\frac{{\dot a}}{{Na}}} )^2  + \frac{{\dot \phi ^I \dot \phi ^I }}{{N^2 }} - \lambda  + M_{Pl}^2 \frac{k}{{a^2 }})
\end{equation}
Where the term with $k$ is the spatial curvature which is positive, negative or zero depending on whether the spatial geometry of the universe is positively curved, negatively curved or flat. In going from (2.1) to (2.2) we have rescaled the Planck mass to keep the appearance of formulas simple. Defining the volume as $V=a^3$ we have:

\[
S = \frac{1}{2}\int {dtVN( - M_{Pl}^2 (\frac{{\dot V}}{{NV}}} )^2  + \frac{{\dot \phi ^I \dot \phi ^I }}{{N^2 }} - \lambda  + M_{Pl}^2 \frac{k}{{V^{2/3} }})
\]
Defining $\tilde N = NV$ the action is written:

\[
S = \frac{1}{2}\int {dt\tilde N( - M_{Pl}^2 (\frac{{\dot V}}{{\tilde N}}} )^2  + V^2 \frac{{\dot \phi ^I \dot \phi ^I }}{{\tilde N^2 }} - \lambda  + M_{Pl}^2 \frac{k}{{V^{2/3} }})
\]
We see that this form of the action is equivalent to a particle moving in a background $(V,\phi^I)$ space. From the form of the action we can define a metric in this $(V,\phi^I)$ space as:

\[
\delta s^2  =  - M_{Pl}^2 dV^2  + V^2 d\phi ^I d\phi ^I 
\]
Now after variation with respect to $\tilde N$ and setting $\pi_V = \frac{1}{\tilde N}M_{Pl}^2 \dot V$, and $\pi^I = \frac{1}{\tilde N}V^2 \phi^I$ one obtains the constraint:

\[
 - \frac{1}{{M_{Pl}^2 }}\pi _V^2  + \frac{1}{{V^2 }}\pi _\phi ^I \pi _\phi ^I  - M_{Pl}^2 \frac{k}{{V^{2/3} }} + \lambda = 0
\]
Turning the canonical momenta into operators we write the Wheeler-DeWitt equation as:

\[
( -\frac{1}{{M_{Pl}^2 }}\frac{\partial }{{\partial V}}\frac{\partial }{{\partial V}} + \frac{1}{{V^2 }}\frac{\partial }{{\partial \phi ^I }}\frac{\partial }{{\partial \phi ^I }}  + M_{Pl}^2 \frac{k}{{V^{2/3} }})\Phi  = \lambda \Phi
\]
where $\Phi$ is the WDW wave function or wave function of the universe.
The important point for this paper is that that the cosmological constant term acts like a $Mass^2$ term in the equation. This was used in \cite{McGuigan:1989yb} to describe the decay of a universe with a large value of $\lambda$ to a universe with a small value of $\lambda$ in a process analogous to particle decay. As we shall see the cosmological constant seesaw mechanism in quantum cosmology is somewhat different in that one not only has the possibility of decay but also oscillation and mixing of universes through off diagonal terms in the Wheeler -De Witt equation.

Another important point is that there is no time in the WDW equation until further gauge fixing is chosen. It is only that the WDW equation is of the Klein Gordon type and $d/dV$ plays the role that $d/dt$ plays in the Klein Gordon equation. Asymptotically one can decompose the metric into transverse traceless parts and use coordinate choices to define a notion of time. In particular in 2+1 dimensions one can choose a notion of time so that one can define a equation first order in time derivatives by taking a Dirac square root \cite{Carlip:1991ij}.  

Perhaps the most straightforward approach to the interpretation of the WDW equation is to look at the form of the solutions at large volume $V$and see how they relate to a semiclassical description around classical cosmological solutions. For a particular choice of operator ordering the solution to the WDW equation for $k=0$ is given by Hankel functions of the form:

\[
\Phi \sim H_{iM_{Pl} \left| {\pi _\phi  } \right|}^{(2)} (\sqrt {\lambda  } M_{Pl} V)
\]
where we used a separation of variables 

\[
\Phi (V,\phi ^I ) = \Phi _{\pi _\phi  } (V)\exp(i\phi ^I \pi _\phi ^I )
\]
Other choices of operator  ordering only introduce a volume dependence prefactor which doesn't effect our discussion. For $k$ not equal to zero one can use a WKB type solution of the form:

\[
\Phi \sim \exp( - iM_{Pl} \int\limits_{V_0 }^V {dV'\sqrt {\lambda  + \pi _\phi ^I \pi _\phi ^I V'^{ - 2}  - M_{Pl}^2 kV'^{ - 2/3} } } )
\]
For either of these solutions the large $V$ behavior is given by:

\begin{equation}
 \Phi  \sim \exp( - iM_{Pl} \sqrt {\lambda  } V) +  \ldots   
\end{equation}
Thus we say that a solution to to the WDW equation describes a universe with cosmological constant $\lambda$ if it's large $V$ behavior is given by (2.3).

\subsection{Universe mixing and coupled WDW equations}

In the Wheeler-DeWitt approach to quantum gravity each universe obeys an equation which can be interpreted as a field equation in $(V,\phi)$ space. This gives a simple way to introduce an interaction between universes by coupling these equations. Alternatively one can use a path integral approach where one evolves the geometry to a conifold space which serves as an intermediary space and then evolve the geometry to a different topology. This can be made more explicit in UV complete models such as 1+1 gravity coupled to matter or 2+1 gravity without matter. 

In this paper we expect that our results can be given a Ultraviolet (UV) completion  through a finite theory of gravity like string theory. We define the term UV complete as any theory in which energy can be taken to infinity (or length scales taken to zero) without any modification of the theory, that is without the introduction of any new physics at those scales. One can even go further and define strongly UV complete theories as those which are background independent, weakly UV complete theories which are background dependent, and kinematically UV complete theories which are finite in certain kinematic limits. In string models (which are not yet known to be strongly UV complete because of background dependence) one can follow a time dependent  sigma model to a conifold and then out to another geometry to describe a mild form of topology change. Also recently the Wheeler-DeWitt approach has been placed on somewhat firmer ground by calculating the wave function of the universe exactly in certain string models \cite{Ooguri:2005vr}. For M-theory which is known to be finite in certain kinematic limits instead of the sigma model one can use a dual Matrix theory to define a big bang cosmology \cite{Craps:2005wd}. 

Here we take the approach of coupling two WDW equations to allow one universe to transform into another. In a later section we will discuss how the results we obtain can be cast into more exact treatments in UV complete models.
We begin by writing two WDW equations for two universes one with cosmological constant $\lambda_1$ and one with a  cosmological constant parametrized by $\tilde \lambda_2 = \lambda_1 + \lambda_2$ without coupling. 

\[
\begin{array}{l}
 (-\frac{1}{{M_{Pl}^2 }}\frac{\partial }{{\partial V}}\frac{\partial }{{\partial V}} + \frac{1}{{V^2 }}\frac{\partial }{{\partial \phi ^I }}\frac{\partial }{{\partial \phi ^I }}  + M_{Pl}^2 \frac{k}{{V^{2/3} }} -\lambda_1)\Phi _1    = 0 \\ 
 (-\frac{1}{{M_{Pl}^2 }}\frac{\partial }{{\partial V}}\frac{\partial }{{\partial V}} + \frac{1}{{V^2 }}\frac{\partial }{{\partial \phi ^I }}\frac{\partial }{{\partial \phi ^I }} + \ + M_{Pl}^2 \frac{k}{{V^{2/3} }}-(\lambda _1  + \lambda _2 ))\Phi _2  = 0 \\ 
 \end{array}
\]
Writing the second cosmological constant as the sum of two quantities $\tilde \lambda_2 = \lambda _1  + \lambda _2$ is for convenience when we discuss the inclusion of mixing. 
As in our discussion above we can interpret this equation as of the Klein-Gordon form with $Mass^2$ matrix:

\[
M^2  =  \left( {\begin{array}{*{20}c}
   {\lambda _1 } & {0}   \\
   {0} & {\lambda _1  + \lambda _2 }  \\
\end{array}} \right)
\]

 Now we introduce a coupling between these two equations. The motivation is that topology change can be expected in any quantum theory of gravity. When present it effectively allows universes of one type or topology to connect or turn into another. Allowing for different values of the cosmological constant associated with different topologies is natural in string inspired models where different compactifications lead to different supersymmetry breaking and different values of $\lambda$. We will discuss some of these models in later sections. The coupling between ground state wave functions is also seen in $1+1$ solvable gravity models describing a midisuperspace topology change between $I=[0,1]$ and $S^1$ spatial topology. In string theory these give a world sheet description of the interaction between open and closed string states.

The coupling between the two WDW equations is taken as: 

\[
\begin{array}{l}
 (-\frac{1}{{M_{Pl}^2 }}\frac{\partial }{{\partial V}}\frac{\partial }{{\partial V}} + \frac{1}{{V^2 }}\frac{\partial }{{\partial \phi ^I }}\frac{\partial }{{\partial \phi ^I }}  - M_{Pl}^2 \frac{k}{{V^{2/3} }}+ \lambda _1  )\Phi _1  + \sqrt {\lambda _1 \lambda _2 } \Phi _2  = 0 \\ 
 (-\frac{1}{{M_{Pl}^2 }}\frac{\partial }{{\partial V}}\frac{\partial }{{\partial V}} + \frac{1}{{V^2 }}\frac{\partial }{{\partial \phi ^I }}\frac{\partial }{{\partial \phi ^I }}  - M_{Pl}^2 \frac{k}{{V^{2/3} }}+ \lambda _1  + \lambda _2  )\Phi _2  + \sqrt {\lambda _1 \lambda _2 } \Phi _1  = 0 \\ 
 \end{array}
\]
So that the effective $M^2$ matrix is modified to :

\[
M^2  = \left( {\begin{array}{*{20}c}
   0 & {\sqrt {\lambda _1 } }  \\
   {\sqrt {\lambda _1 } } & {\sqrt {\lambda _2 } }  \\
\end{array}} \right)\left( {\begin{array}{*{20}c}
   0 & {\sqrt {\lambda _1 } }  \\
   {\sqrt {\lambda _1 } } & {\sqrt {\lambda _2 } }  \\
\end{array}} \right) = \left( {\begin{array}{*{20}c}
   {\lambda _1 } & {\sqrt {\lambda _1 \lambda _2 } }  \\
   {\sqrt {\lambda _1 \lambda _2 } } & {\lambda _1  + \lambda _2 }  \\
\end{array}} \right)
\]
This form is motivated by the ordinary seesaw mechanism and is somewhat more natural from point of view of taking a Dirac square root and also in supergavity where it is the $M$ matrix which plays the major role.

The coupled set of Wheeler-DeWitt equations can be derived from the Lagrangian:
\[
L = \left( {\begin{array}{*{20}c}
   {\Phi _1^* } & {\Phi _2^* }  \\
\end{array}} \right)\left( {\begin{array}{*{20}c}
   \Delta  & 0  \\
   0 & \Delta   \\
\end{array}} \right)\left( {\begin{array}{*{20}c}
   {\Phi _1 }  \\
   {\Phi _2 }  \\
\end{array}} \right) + \left( {\begin{array}{*{20}c}
   {\Phi _1^* } & {\Phi _2^* }  \\
\end{array}} \right)\left( {\begin{array}{*{20}c}
   {\lambda _1 } & {\sqrt {\lambda _1 \lambda _2 } }  \\
   {\sqrt {\lambda _1 \lambda _2 } } & {\lambda _1  + \lambda _2 }  \\
\end{array}} \right)\left( {\begin{array}{*{20}c}
   {\Phi _1 }  \\
   {\Phi _2 }  \\
\end{array}} \right)
\]
Where we have defined the operator:

\begin{equation}
\Delta  = \frac{1}{{M_{Pl}^2 }}\frac{\partial }{{\partial V}}\frac{\partial }{{\partial V}} - \frac{1}{{V^2 }}\frac{\partial }{{\partial \phi ^I }}\frac{\partial }{{\partial \phi ^I }} - M_{Pl}^2 \frac{k}{{V^{2/3} }}
\end{equation}
As with the usual seesaw mechanism we can diagonalize the coupled equations  by computing the eigenvalues and eigenvectors of the matrix $M$. The eigenvalues are given by:

\begin{equation}
\begin{array}{l}
 \lambda _ -   = \lambda _1  + \frac{1}{2}\lambda _2  - \sqrt {(\lambda _1  + \frac{1}{2}\lambda _2 )^2  - \lambda _1^2 }  \\ 
 \lambda _ +   = \lambda _1  + \frac{1}{2}\lambda _2  + \sqrt {(\lambda _1  + \frac{1}{2}\lambda _2 )^2  - \lambda _1^2 }  \\ 
 \end{array}
\end{equation}

These eigenvalues satisfy the relation:

\[
\lambda _ -  \lambda _ +   = \lambda _1^2 
\]
Expanding out the relation when $\lambda _2$ is much greater than $\lambda_1$ we have:

\begin{equation}
\begin{array}{l}
 \lambda _ -   = \frac{{\lambda _1^2 }}{{2\lambda _1  + \lambda _2 }} +  \ldots  \\ 
 \lambda _ +   = 2\lambda _1  + \lambda _2  +  \ldots  \\ 
 \end{array}
\end{equation}
It is sometimes useful to use the ${\tilde{\lambda}}_2$ parametrization. In this case

\[
\lambda _ \pm   = \frac{{\lambda _1  + \tilde \lambda _2 }}{2} \pm \frac{{\sqrt {(\lambda _1  + \tilde \lambda _2 )^2  - 4\lambda _1^2 } }}{2}
\]
which becomes for $\tilde \lambda _2$ much greater than $\lambda_1$:

\[
\begin{array}{l}
 \lambda _ -   = \frac{{\lambda _1^2 }}{{\lambda _1  + \tilde \lambda _2 }} +  \ldots  \\ 
 \lambda _ +   = \lambda _1  + \tilde \lambda _2  +  \ldots  \\ 
 \end{array}
\]

We can expand the Wheeler-DeWitt wave functions in an eigenbasis of the matrix to obtain:

\[
\left( {\begin{array}{*{20}c}
   {\Phi _1 }  \\
   {\Phi _2 }  \\
\end{array}} \right) = \Phi _ -  \frac{1}{{1 + \lambda _ -  /\lambda _1 }}\left( {\begin{array}{*{20}c}
   1  \\
   { - \sqrt {\lambda _ -  /\lambda _1 } }  \\
\end{array}} \right) + \Phi _ -  \frac{1}{{1 + \lambda _ +  /\lambda _1 }}\left( {\begin{array}{*{20}c}
   1  \\
   {\sqrt {\lambda _ +  /\lambda _1 } }  \\
\end{array}} \right)
\]
Using the new wave functions the Lagrangian becomes:

\[
L = \left( {\begin{array}{*{20}c}
   {\Phi _ - ^* } & {\Phi _ + ^* }  \\
\end{array}} \right)\left( {\begin{array}{*{20}c}
   \Delta  & 0  \\
   0 & \Delta   \\
\end{array}} \right)\left( {\begin{array}{*{20}c}
   {\Phi _ -  }  \\
   {\Phi _ +  }  \\
\end{array}} \right) + \left( {\begin{array}{*{20}c}
   {\Phi _ - ^* } & {\Phi _ + ^* }  \\
\end{array}} \right)\left( {\begin{array}{*{20}c}
   {\lambda _ -  } & 0  \\
   0 & {\lambda _ +  }  \\
\end{array}} \right)\left( {\begin{array}{*{20}c}
   {\Phi _ -  }  \\
   {\Phi _ +  }  \\
\end{array}} \right)
\]
The decoupled Wheeler-Dewitt equations  derived from this Lagrangian are then:

\[
\begin{array}{l}
 (\frac{1}{{M_{Pl}^2 }}\frac{\partial }{{\partial V}}\frac{\partial }{{\partial V}} - \frac{1}{{V^2 }}\frac{\partial }{{\partial \phi ^I }}\frac{\partial }{{\partial \phi ^I }} +   - M_{Pl}^2 \frac{k}{{V^{2/3}}}+ \lambda _ - )\Phi _ -   = 0 \\ 
 (\frac{1}{{M_{Pl}^2 }}\frac{\partial }{{\partial V}}\frac{\partial }{{\partial V}} - \frac{1}{{V^2 }}\frac{\partial }{{\partial \phi ^I }}\frac{\partial }{{\partial \phi ^I }}   - M_{Pl}^2 \frac{k}{{V^{2/3} }} +\lambda_+)\Phi _ +   = 0 \\ 
 \end{array}
\]
Using the expressions for the eigenvalues in (2.4) and setting

\[
\begin{array}{l}
 \lambda _1  = (10TeV)^4  \\ 
 \lambda _2  = M_{Pl}^4  \\ 
 \end{array}
\]
we obtain the desired expression:

\[
\begin{array}{l}
 \lambda _ -   = \frac{{(10TeV)^8 }}{{M_{Pl}^4 }} +  \ldots  \\ 
 \lambda _ +   = M_{Pl}^4  +  \ldots  \\ 
 \end{array}
\]
The wave functions $\Phi_-$ and $\Phi_+$ are then given by:

\[
\begin{array}{l}
 \Phi _- \sim H_{iM_{Pl} \left| {\pi _\phi  } \right|}^{(2)} (\sqrt {\lambda _ -  } M_{Pl} V)  \\ 
 \Phi _+ \sim  H_{iM_{Pl} \left| {\pi _\phi  } \right|}^{(2)} (\sqrt {\lambda _ +  } M_{Pl} V) \\ 
 \end{array}
\]

Here we discuss the $k=0$ case. For $k$ not  equal to zero we can use WKB solutions.
Expressing the original fields $\Phi_1$ and $\Phi_2$ in terms of $\Phi_-$ and $\Phi_+$ we have the solution to the original coupled set of equations as:

\[
\begin{array}{l}
 \Phi _1 \sim H_{iM_{Pl} \left| {\pi _\phi  } \right|}^{(2)} (\sqrt {\lambda _ -  } M_{Pl} V) + \sqrt {\frac{{\lambda _1 }}{{\lambda _2 }}} H_{iM_{Pl} \left| {\pi _\phi  } \right|}^{(2)} (\sqrt {\lambda _ +  } M_{Pl} V) \\ 
 \Phi _2 \sim  - \sqrt {\frac{{\lambda _1 }}{{\lambda _2 }}} H_{iM_{Pl} \left| {\pi _\phi  } \right|}^{(2)} (\sqrt {\lambda _ -  } M_{Pl} V) + H_{iM_{Pl} \left| {\pi _\phi  } \right|}^{(2)} (\sqrt {\lambda _ +  } M_{Pl} V) \\ 
 \end{array}
\]
The behavior of these wave functions at large $V$ are given by 
 
\[
\begin{array}{l}
 \Phi _1 \sim exp( - iM_{Pl} \sqrt {\lambda _ -  } V) +  \ldots  \\ 
 \Phi _2 \sim exp( - iM_{Pl} \sqrt {\lambda _ +  } V) +  \ldots  \\ 
 \end{array}
\]
plus a small mixing term. Thus we see that they describe predominantly a classical cosmology with cosmological constant $\lambda_-$ and $\lambda_+$ respectively. The cosmological constant seesaw mechanism is then the identification of our universe as the dominant piece in $\Phi_1$ which describes at large $V$ a cosmology evolving with cosmological constant  $\lambda _ -   = \frac{{(10TeV)^8 }}{{M_P^4 }} +  \ldots$.

Let's review what we did. We identified the cosmological constant as a $Mass^2$ type term in the Wheeler-DeWitt equation. We coupled two such equations together through a mixing term and formed a $2\times 2$ mass matrix whose form was inspired from the ordinary seesaw mechanism. We solved this coupled set of equations and looking at the large $V$ behavior of the solutions and found a phenomenologically viable expression for the cosmological constant in terms of two constants describing the $Mass^2$ matrix. The size of these constants is  of order of the vacuum energy before mixing which we take to be $(10TeV)^4$ in one universe and $M_{Pl}^4$ in another. We discuss a common mechanism to explain the appearance of the $10TeV$ scale in later sections.

One interpretation of the coupling of Wheeler-DeWitt functions is that it originates from topology changing effects. Topology change seems to be inevitable in quantum gravity. To treat topology change properly is a very complicated calculation using today's mathematical tools. To place this in context  even a simple transition from spatial topology of $I$ to $S^1$ where $I$ is the unit interval is at least as complicated as the transition from open to closed strings in String Field theory. Indeed  in this case the topology change leads to a coupling between open and closed  $A_\mu$ and  $B_{\mu\nu}$ wave functions. This has a physical effect of the  $B_{\mu\nu}$ field eating the  $A_\mu$ field and becoming massive and modifies it's wave equation. The effect we are studying is similar to this case except we a dealing with topology change in spacetime, wave equations in $(V, \phi^I)$ space and the cosmological constant instead of masses. We shall find an even closer  analogy to the neutrino seesaw in supersymmetric theories where it is the $Mass$ instead of $Mass^2$ matrix which appears in the constraint equations.

\section{String Inspired Gravity} 

In the next three sections we apply the techniques of the previous section to various gravity matter systems of current interest.

Contrary to popular belief string theory does make some model independent predictions. In particular the string dilaton and axion are present in almost all string models. In this case the Einstein-Hilbert action is of the form:

\[
S = \frac{1}{2}\int {d^4 x(M_{Pl}^2 \sqrt { - g} } R - \sqrt { - g} \lambda  - \partial \phi \partial \phi  - e^{2\phi } \partial \chi \partial \chi )
\]
Where $\phi$ is related to the dilaton and $\chi$ to the axion and we have Weyl transformed to the Einstein frame using the notation of \cite{Maharana:2004qs}. The axion is related to the antisymmetric field through the relation:

\[
H^{\mu \nu \rho }  = e^{2\phi } \varepsilon ^{\mu \nu \rho \sigma } \partial _\sigma  \chi 
\]
Taking the ansatz $\phi = \phi(t)$ and $\chi = \chi(t)$ and defining:

\[
\begin{array}{l}
 m_1 = \chi  \\ 
 m_2 = e^{ - \phi }  \\ 
 \end{array}
\]
the action becomes:
\[
S = \frac{1}{2}\int {dt\tilde N( - M_{Pl}^2 (\frac{{\dot V}}{{\tilde N}}} )^2  + M_{Pl}^2 V^2 \frac{{\dot m_1^2 + \dot m_2^2 }}{{m_2^2\tilde N^2 }} - \lambda  + M_{Pl}^2 \frac{k}{{V^{2/3} }})
\]
This  action is equivalent to that of a particle with $Mass^2=\lambda$ with background metric in $(V,m_1,m_2)$ space given by:

\[
\delta s^2  =  - M_{Pl}^2 dV^2  + M_{Pl}^2 V^2 \frac{1}{{m_2^2 }}(dm_1^2  + dm_2^2 )
\]
The constraint derived from the above action is:

\[
 - \frac{1}{{M_{Pl}^2 }}\pi _V^2  + \frac{m_2^2}{{M_{Pl}^2V^2 }}(\pi _1 ^2+ \pi _2 ^2) - M_{Pl}^2 \frac{k}{{V^{2/3} }} + \lambda = 0
\]
 where $\pi_1$ and $\pi_2$ are the canonical momentum associated with $m_1$ and $m_2$. As before treating this as a operator equation we write the WDW equation as:

\[
( \frac{1}{{M_{Pl}^2 }}\frac{\partial^2}{\partial V^2}  -  \frac{m_2^2}{{M_{Pl}^2V^2 }}(\frac{\partial^2}{\partial m_1^2}+  \frac{\partial^2}{\partial m_1^2}) - M_{Pl}^2 \frac{k}{{V^{2/3} }} + \lambda)\Phi = 0
\]
To solve this equation one can again try separation of variables:
\[
\Phi(v,m_1,m_2) = \Phi_s(V)f_s(m_1,m_2)
\]
where $f_s$ obeys:
\[
m_2^2(\frac{\partial^2}{\partial m_1^2}+  \frac{\partial^2}{\partial m_1^2})f_s = s(s-1)f_s 
\]
and $\Phi_s$ satisfies the equation:

\[
( \frac{1}{{M_{Pl}^2 }}\frac{\partial^2}{\partial V^2}  +  \frac{s(1-s)}{{M_{Pl}^2V^2 }} - M_{Pl}^2 \frac{k}{{V^{2/3} }} + \lambda)\Phi_s = 0
\]
Analysis of the first equation \cite{Maharana:2004qs} indicates that the solutions of cosmological interest have $s=\frac{1}{2}+i\nu$ with $\nu > 0$. For a certain choice of operator ordering and $k=0$ the second equation has the solution of the form:

\[
\Phi_s(V) \sim H_{i\sqrt{\nu^2+\frac{1}{4}}}^{(2)} (\sqrt{\lambda}VM_{Pl})
\]
For $k$ not equal to zero the solution is more complicated but we can use  a WKB type solution:
\[
\Phi_s(V) \sim \exp( - iM_{Pl} \int\limits_{V_0 }^V {dV'\sqrt {\lambda  + s(1-s)M_{Pl}^{-2}V'^{ - 2}  - M_{Pl}^2 kV'^{ - 2/3} } } )
\]

The cosmological seesaw proceeds as in the previous section. One introduces a pair of coupled WDW equations of the form:

\[
\begin{array}{l}
 (\frac{1}{{M_{Pl}^2 }}\frac{\partial }{{\partial V}}\frac{\partial }{{\partial V}} + \frac{s(1-s)}{{M_{Pl}^2V^2 }} - M_P^2 \frac{k}{{V^{2/3} }}+ \lambda _1  )\Phi _1  + \sqrt {\lambda _1 \lambda _2 } \Phi _2  = 0 \\ 
 (-\frac{1}{{M_{Pl}^2 }}\frac{\partial }{{\partial V}}\frac{\partial }{{\partial V}} + \frac{s(1-s)}{{V^2 }} - M_{Pl}^2 \frac{k}{{V^{2/3} }}+ \lambda _1  + \lambda _2  )\Phi _2  + \sqrt {\lambda _1 \lambda _2 } \Phi _1  = 0 \\ 
 \end{array}
\]
and proceeds to solve these coupled equations. In the above we have suppressed the $s$ index. The main difference from the previous case is the parameter $\sqrt{\nu^2+ \frac{1}{4}}$ in the Hankel function instead of $\sqrt{\pi_{\phi}^I\pi_{\phi}^I}$ so that: 

\[
\begin{array}{l}
 \Phi _1 \sim H_{i\sqrt{\nu^2+ \frac{1}{4}}}^{(2)} (\sqrt {\lambda _ -  } M_{Pl} V) + \sqrt {\frac{{\lambda _1 }}{{\lambda _2 }}} H_{i\sqrt{\nu^2+ \frac{1}{4}}}^{(2)} (\sqrt {\lambda _ +  } M_{Pl} V) \\ 
 \Phi _2 \sim  - \sqrt {\frac{{\lambda _1 }}{{\lambda _2 }}} H_{i\sqrt{\nu^2+ \frac{1}{4}}}^{(2)}(\sqrt {\lambda _ -  } M_{Pl} V) + H_{i\sqrt{\nu^2+ \frac{1}{4}}}^{(2)}(\sqrt {\lambda _ +  } M_{Pl} V) \\ 
 \end{array}
\]
For large $V$ the cosmological constant term dominates and this difference doesn't change the large $V$ behavior of the solutions which again go as :
 
\[
\begin{array}{l}
 \Phi _1 \sim \exp( - iM_{Pl} \sqrt {\lambda _ -  } V) +  \ldots  \\ 
 \Phi _2 \sim \exp( - iM_{Pl} \sqrt {\lambda _ +  } V) +  \ldots  \\ 
 \end{array}
\]
One identifies our universe with the the portion of the wave function  with an effective small cosmological constant $\lambda_-$.

The system of axion/dilaton gravity is formally similar to the case of 2+1 quantum gravity with spatial topology $T^2$ except in the case of 2+1 gravity there are no gravitons and the minisuperspace description is exact in certain formulations. We shall return to exactly soluble systems in a later section. For here we note that like the system of 2+1 gravity \cite{Carlip:1991ij} one can construct a Dirac square root of the coupled WDW equation by writing:

\[
\begin{array}{l}
 \frac{1}{{M_{Pl} V}}\left( {\begin{array}{*{20}c}
   {V\pi _V } & {im_2 (\pi _1  + i\pi _2 )}  \\
   {im_2 (\pi _1  - i\pi _2 )} & { - V\pi _V }  \\
\end{array}} \right)\Psi _1  = \sqrt {\lambda _1 } \Psi _2  \\ 
 \frac{1}{{M_{Pl} V}}\left( {\begin{array}{*{20}c}
   {V\pi _V } & {im_2 (\pi _1  + i\pi _2 )}  \\
   {im_2 (\pi _1  - i\pi _2 )} & { - V\pi _V }  \\
\end{array}} \right)\Psi _2  = \sqrt {\lambda _1 } \Psi _1  + \sqrt {\lambda _2 } \Psi _2  \\ 
 \end{array}
\]
where $\Psi_1$ and $\Psi_2$ are two component wave functions describing separate universes. To take into account operator ordering one can replace $im_2 (\pi _1  + i\pi _2 )$ with $L_1  =  - m_2 (\frac{\partial }{{\partial m_1 }} + i\frac{\partial }{{\partial m_1 }}) - 1$ and  $im_2 (\pi _1  - i\pi _2 )$ with $K_0  = m_2 (\frac{\partial }{{\partial m_1 }} - i\frac{\partial }{{\partial m_1 }})$ as in \cite{Carlip:1991ij}. It's clear that in this form it is the matrix:

\[
M = \left( {\begin{array}{*{20}c}
   0 & {\sqrt {\lambda _1 } }  \\
   {\sqrt {\lambda _1 } } & {\sqrt {\lambda _2 } }  \\
\end{array}} \right)
\]
which appears. So the Dirac square root method is quite similar to the ordinary seesaw mechanism. 

One interesting feature of  string inspired actions is  S-duality symmetry.  S-duality is given by:

\[
m_1+im_2 \rightarrow \frac{a(m_1+im_2)+b}{c(m_1+im_2)+d}
\]
where $a,b,c,d$ are integers such that $ad-bc=1$. This symmetry relates strong coupling to weak coupling. Also in any realistic application a potential $U(m_1,m_2)$ should be introduced. To preserve S-duality $U(m_1,m_2)$ and the wave functions should transform as modular forms. For $k=0$ one can also impose a T-duality which relates $V$ to $M_{Pl}^{-6}/V$ \cite{McGuigan:1990pi}\cite{Smith:1991up}\cite{Gasperini:1991ak}. This is particularly interesting with respect to the cosmological constant seesaw because it relates the large $V$ behavior of the wave function to it's small $V$ behavior. In quantum cosmology the large $V$ behavior is dominated by the cosmological constant. By contrast the small $V$ behavior is dominated by short distance physics. The T-duality symmetry relates these limits.

Another interesting feature of string inspired gravity is that  $\lambda$ can be computed in string models. Typically one gets values for $\lambda$ that are of order of the supersymmetry breaking scale (supersymmetric models) or of order of the planck scale (nonsupersymmetric models) and this fits  the input parameters $\lambda_1$ and $\lambda_2$ of the seesaw model quite well.

\section{M-theory Inspired Gravity}

There are at least two separate approaches to M-theory inspired gravity. In one approach called the Bousso-Polchinski model \cite{Bousso:2000xa} one compactifies the 11d theory to four dimensions on a seven manifold and turns on several (possibly hundreds) of four form fluxes which adjust a bare cosmological constant. In another approach which we will call the Horava-Witten model \cite{Horava:1996ma} one compactifies first to five dimensions on a six manifold and then treats ordinary matter as fixed to four dimensional boundaries which can interact gravitationally through bulk exchange of five dimensional gravitons and graviphotons. In this context one has also an solution to the Hierarchy problem discussed in \cite{Randall:1999ee}.

We shall discuss these models separately and seek to place them in the context of the cosmological constant seesaw.

\subsection{Bousso-Polchinski Model}

The Bousso-Polchinski model can be described by the low energy action:

\[
S = \frac{1}{2}\int {d^4 x(M_{Pl}^2 \sqrt { - g} } R - \sqrt { - g} \lambda  - \frac{1}{{24}}F^{(I)\mu \nu \rho \sigma } F_{\mu \nu \rho \sigma }^{(I)} )
\]
where $F^I =dA^I$ are a set of four forms and $\lambda$ is a bare cosmological constant.
Following \cite{Duncan:1990fr} we introduce a three form potential ansatz:

\[
A_{ijk}^{(I)} (t) = \varepsilon _{ijk} \varphi ^I (t)
\]
In the Bouso-Polchinski model the four fluxes are quantized. This translates in minisuperspace to periodicity in $\varphi^I \rightarrow \varphi^I+2\pi R^{I2}$ where
$R^{I2}$ are parameters of dimension of area which characterize the flux.

Using the above ansatz the action becomes:

\[
S = \frac{1}{2}\int {dtNa^3 ( - 9M_{Pl}^2 \frac{{\dot a^2 }}{{N^2 a^2 }}}  - \lambda  + \frac{{\dot \varphi ^I \dot \varphi ^I }}{{N^2 a^6 }} + M_{Pl}^2 \frac{k}{{a^2 }})
\]
Defining $\tilde N = NV$ the action is written as:

\[
S = \frac{1}{2}\int {dt\tilde N( - M_{Pl}^2 (\frac{{\dot V}}{{\tilde N}}} )^2  + \frac{{\dot \varphi ^I \dot \varphi ^I }}{{\tilde N^2 }} - \lambda  + M_{Pl}^2 \frac{k}{{V^{2/3} }})
\]
In this case we can define a metric in this $(V,\phi^I)$ space as

\[
\delta s^2  =  - M_P^2 dV^2  +  d\varphi ^I d\varphi ^I 
\]
Now after variation with respect to $\tilde N$ and setting $\pi_V = M_{Pl}^2 \dot V/\tilde N$, and $\pi_{\varphi}^I = \dot{\varphi^I}/\tilde N $ one obtains the constraint:

\[
 - \frac{1}{{M_{Pl}^2 }}\pi _V^2  + \pi _\varphi ^I \pi _\varphi ^I  - M_{Pl}^2 \frac{k}{{V^{2/3} }} + \lambda = 0
\]
The WDW equation is then:

\[
( \frac{1}{{M_{Pl}^2 }}\frac{\partial^2 }{{\partial V^2}} - \frac{\partial }{{\partial \varphi ^I }}\frac{\partial }{{\partial \varphi ^I }}  - M_{Pl}^2 \frac{k}{{V^{2/3} }}  +\lambda) \Phi = 0
\]
Using the separation of variables:

\[
\Phi(V,\varphi^I) = \Phi_n(V) \exp({in^I\varphi^I/R^{I2}})
\]
it is straightforward to solve this equation. For $k=0$ we have the simple solution:

\[
\Phi(V,\varphi^I) = \exp({-iVM_{Pl} \sqrt{\frac{n^{I2}}{R^{I4}}+\lambda}}) \exp({in^I\varphi^I/R^{I2}})
\]
For $k$ not equal to zero we can follow Duncan and Jensen \cite{Duncan:1990fr}   to define:

\[
z = \left( {\frac{{3VM_{Pl} }}{{2(\frac{{n^{I2} }}{{R^{I4} }} + \lambda )}}} \right)^{2/3} (\frac{{n^{I2} }}{{R^{I4} }} + \lambda  - k\frac{{M_{Pl}^2 }}{{V^{2/3} }})
\]
The solution for a certain choice of operator ordering can be written in terms of Airy functions as:
\[
\Phi _n (V) = Ai(z) - iBi(z)
\]
Other choices of operator ordering will not modify the large $V$ behavior of the wave function which we use in our analysis. Using the asymptotic expression for Airy functions we have:

\[
\Phi _n (V) \sim \exp ( - i\frac{{VM_{Pl} }}{{(\frac{{n^{I2} }}{{R^{I4} }} + \lambda )}}(\frac{{n^{I2} }}{{R^{I4} }} + \lambda  - k\frac{{M_{Pl}^2 }}{{V^{2/3} }})^{3/2} )
\]
So that we have the same large $V$ behavior as $k=0$:

\[
\Phi _n (V) \sim \exp( { - iVM_{Pl} \sqrt {\frac{{n^{I2} }}{{R^{I4} }} + \lambda } })
\]
In this form we see that $\lambda_{nR} =\frac{{n^{I2} }}{{R^{I4} }} + \lambda$ plays the role of an effective cosmological constant.

The cosmological constant seesaw can be implemented by introducing the mixing between two WDW fields of the form:

\[
\begin{array}{l}
 (-\frac{1}{{M_{Pl}^2 }}\frac{\partial }{{\partial V}}\frac{\partial }{{\partial V}} + \frac{1}{{V^2 }}\frac{\partial }{{\partial \phi ^I }}\frac{\partial }{{\partial \phi ^I }}  + M_{Pl}^2 \frac{k}{{V^{2/3} }} -(  \frac{{n^{I2} }}{{R^{I4} }} + \lambda_1)  )\Phi _1    = \sqrt{   \lambda_{1nR} \lambda_{2mL}}\Phi2                \\ 
 (-\frac{1}{{M_{Pl}^2 }}\frac{\partial }{{\partial V}}\frac{\partial }{{\partial V}} + \frac{1}{{V^2 }}\frac{\partial }{{\partial \phi ^I }}\frac{\partial }{{\partial \phi ^I }} + \ + M_{Pl}^2 \frac{k}{{V^{2/3} }}-( \frac{{m^{I2} }}{{L^{I4} }}+ \tilde\lambda _2 ))\Phi _2  
= \sqrt{   \lambda_{1nR} \lambda_{2mL}}\Phi1
  \\ 
 \end{array}
\]
Here $(n^I/R^{I2},\lambda_1)$ are the flux and cosmological constant in one universe.\\
$(m^I/L^{I2},\tilde\lambda_2)$ are the flux and cosmological constant in the other universe. As before $\tilde \lambda_2= \lambda_1+\lambda_2$.
The solutions to the coupled equations are again of the form:

\[
\begin{array}{l}
 \Phi _1 \sim exp( - iM_{Pl} \sqrt {\lambda _ -  } V) +  \ldots  \\ 
 \Phi _2 \sim exp( - iM_{Pl} \sqrt {\lambda _ +  } V) +  \ldots  \\ 
 \end{array}
\]
where in this case:
\[
\lambda _ \pm   = \frac{{\lambda _1  + \frac{{n^{I2} }}{{R^{I4} }} + \tilde \lambda _2  + \frac{{m^{I2} }}{{L^{I4} }}}}{2} \pm \frac{{\sqrt {(\lambda _1  + \frac{{n^{I2} }}{{R^{I4} }} + \tilde \lambda _2  + \frac{{m^{I2} }}{{L^{I4} }})^2  - 4(\lambda _1  + \frac{{n^{I2} }}{{R^{I4} }}} )^2 }}{2}
\]
When the vacuum energy before mixing in the universe described by $\Phi_2$  is much greater than that the vacuum energy in the universe described by $\Phi_1$ we have the approximate formula: 

\begin{equation}
\begin{array}{l}
 \lambda _ -   = \frac{{(\lambda _1  + \frac{{n^{I2} }}{{R^{I4} }})^2 }}{{\lambda _1  + \frac{{n^{I2} }}{{R^{I4} }} + \tilde \lambda _2  + \frac{{m^{I2} }}{{L^{I4} }}}} +  \ldots  \\ 
 \lambda _ +   = \lambda _1  + \frac{{n^{I2} }}{{R^{I4} }} + \tilde \lambda _2  + \frac{{m^{I2} }}{{L^{I4} }} +  \ldots  \\ 
 \end{array}
\end{equation}
These expressions give a phenomenologically viable cosmological constant if:
\[
\begin{array}{l}
 \frac{{n^{I2} }}{{R^{I4} }} + \lambda _1  \approx (10TeV)^4  \\ 
 \frac{{m^{I2} }}{{L^{I4} }} + \tilde\lambda _2  \approx (M_{Pl} )^4  \\ 
 \end{array}
\]
The estimates are quite natural if $\Phi_1$ describes a universe with soft supersymmetry breaking and $\Phi_2$ describes a nonsupersymmetric string or hard supersymmetry breaking at the Planck scale.

An interesting special case is when the parameters $\lambda_1$ and $\tilde\lambda_2$ are zero so that the vacuum energy is completely given by the flux terms. Then:
\[
\begin{array}{l}
 \lambda _ -   \approx \frac{{n^4 }}{{m^2 }}\frac{{L^4 }}{{R^8 }} \\ 
 \lambda _ +   \approx \frac{{m^2 }}{{L^4 }} \\ 
 \end{array}
\]
and for the case when the flux strengths are are determined by the supersymmetry breaking scale in each universe:
\[
\begin{array}{l}
 \frac{1}{R} \approx 10TeV \\ 
 \frac{1}{L} \approx M_{Pl}  \\ 
 \end{array}
\]
we also get phenomenologically viable estimate of the vacuum energy.

Returning to the general case we  note for the WDW equation of this model the cosmological constant again becomes an effective $M^2$ term of the form:

\[
M^2 = \frac{n^{I2}}{R^{I4}}+\lambda
\]
The Bousso-Polchinski mechanism of obtaining a small cosmological constant is quite different from the cosmological constant seesaw we just discussed. The Bousso-Polchinski model effectively adjusts the fluxes $n^{I}/R^{I2}$ so that the $M^2$ is small for negative $\lambda$. In this way we see that the Bousso-Polchinski (BP) mechanism is analogous to the Narain-Sarmadi method of removing Tachyons through compactification \cite{Narain:1986gd}. Except in the BP model the target space is $(V,\varphi^I)$ space and the analog of the Tachyon field is a solution to the Wheeler-DeWitt equation.

One difficulty with Kaluza-Klein approaches such as the BP model are  the large moduli spaces that must be fixed or charges and masses could depend on time. The external radii are dynamical but the internal radii need to be held static and this is unnatural and requires fine tuning. If the  anthropic principle is invoked one comes up against the recent observation \cite{Wilczek:2005aj} that a large number of CKM type parameters of the standard model are irrelevant for life but still appear static (for at least as long as we have been measuring them). 

Another approach might be to pursue a noncritical version of M-theory. See for example \cite{McG}\cite{Hor}\cite{Par}. Noncritical string theories are usually defined with linear dilaton backgrounds or anti-DeSitter space and are not thought to be phenomenologically viable. So little is known about noncritical M-theory that it is tempting to speculate that it might fare better than noncritical string theory as a physical model. Certainly this type of theory would be free of the moduli problem as there aren't any extra dimensions in this approach. Even the dilaton is absent being incorporated as a component of the gravitational field. Large gauge groups are also possible, for example the 2d noncritical Heterotic string contains $E_8 \times SO(8)$ or $SO(24)$ gauge theory \cite{McGuigan:1991qp}\cite{Davis:2005qi}\cite{Davis:2005qe}. For the context of this paper the main point is that non-critical string theory or M-theory contains only a handful of fluxes compared to the Kaluza-Klein theories. For example the $0A$ 2d noncritical string contains two fluxes. Thus the Bousso-Polchinski mechanism might not work in such theories because there are too few fluxes and the cosmological seesaw mechanism could prove to be an alternative.

\subsection{Horava-Witten Model}

In the Horava-Witten approach \cite{Horava:1996ma} one compactifies the 11d M-theory on a six dimensional compact manifold $CY_6$ and then compactifies the remaining extra dimension on an interval $I$. This results in ordinary matter being confined to one of  two 4d boundaries. If the 5d metric connecting the 4d boundaries has a trumpet spatial geometry  it is possible to have a planckian vacuum energy and strong supersymmetry breaking on one four boundary and only a vacuum energy from soft supersymmetry breaking on the other. This allows one to address the Hierarchy problem as in \cite{Randall:1999ee}. For the purposes of this paper it allows us to set up Wheeler- DeWitt equations for two universes with vacuum energies of $(10TeV)^4$ and $M_{Pl}^4$ respectively. This is the starting point for the Cosmological Constant seesaw.

The difference between the previous discussion and the Horava-Witten model is the matter stress-energy on the boundaries that modifies Einstein's equation as \cite{Lukas:1997fg}:

\[
M_{Pl}^2(R_{\mu \nu }  - \frac{1}{2}g_{\mu \nu } R )= (T_{\mu \nu }^{m(a)}  - \lambda _B^{(a)} g_{\mu \nu } )\delta (x_5 ) + (T_{\mu \nu }^{m(b)}  -  \lambda _B^{(b)} g_{\mu \nu } )\delta (x_5  - \pi )
\]
Here $T_{\mu \nu }^{m(a,b)} ,\lambda _B^{(a,b)}$ are the matter stress-energy and boundary cosmological constant on boundary $(a)$ and $(b)$ respectively. Specializing this equation to it's $00$ component tells us how to modify the constraints and the WDW equation. Essentially the approach to the hierarchy problem in \cite{Randall:1999ee}  using the trumpet geometry ensures that on one boundary one has a WDW equation with one type of  boundary matter and small $(10TeV)^4$ boundary cosmological constant $\lambda_B^{(a)}$ while on the other boundary one has another type of boundary matter and planckian boundary cosmological constant $\lambda_B^{(b)}$.

It is not clear whether these two boundary cosmological constants are sufficient to set up the cosmological constant seesaw or if one needs to introduce two separate universes with their different boundary matter and boundary cosmological constants. In the latter case the two WDW equations would essentially be of the form:

\[
\begin{array}{l}
 (\Delta_5 +(T_{00}^{m(a1)}  + \lambda _B^{(a1)}  )\delta (x_5 ) + (T_{00}^{m(b1)}  + \lambda _B^{(b1)} )\delta (x_5  - \pi ))\Phi _1  = 0 \\ 
 (\Delta_5 + (T_{00}^{m(a2)}  + \lambda _B^{(a2)} )\delta (x_5 ) + (T_{00}^{m(b2)}  + \lambda _B^{(b2)} )\delta (x_5  - \pi ))\Phi _2  = 0 \\ 
 \end{array}
\]
In this equation $T_{\mu \nu }^{m(a1,b1)} ,\lambda _B^{(a1,b1)} $ and $T_{\mu \nu }^{m(a2,b2)} ,\lambda _B^{(a2,b2)}$ are boundary stress-energies and cosmological constants in universe 1 and 2 respectively. We have defined $\Delta_5$ as the analog of the operator (2.4) for five dimensional gravity. Here the basic approach is that one universe will connect the two boundaries with a trumpet geometry that will yield a cosmological constant on one boundary at the soft supersymmetry breaking scale. In the other universe the boundaries would be connected by a straight cylindrical geometry so that each boundary would have  Planckian vacuum energy. The cosmological seesaw would correspond to the mixing of these two states.

The presence of mixing between the two WDW wavefunctions depends on the topology change in the model. Estimates of spatial topology transitions computed in 2+1 gravity where one has a UV complete theory, indicate the that the transition probability for spatial topology change is small \cite{Carlip:1994tt}. To go further one needs a higher dimensional UV complete theory in which one can calculate. String theory is strictly speaking not UV complete because of background dependence. Thus it is remarkable that an exact calculation of wave functions of the universe and topology change have been computed in certain string models as we discuss in the next section.

\section{Exact Calculations of the Wave Function of the Universe} 

Most of our discussion has taken place using minisuperspace wavefunctions. To obtain more robust results one needs to either go beyond minisuperspace or work in a context where minisuperspace is exact. In certain approaches to 2+1 gravity with spatial topology $T^2$ one can work with exact minisuperspace wave functions \cite{Martinec:1984fs}. Also in certain string models a minisuperspace description was found and exact wave functions of the universe were obtained \cite{Ooguri:2005vr}. We shall discuss these approaches separately in relation to the cosmological constant seesaw mechanism.

\subsection{2+1 Dimensional Gravity}

For 2+1 gravity without matter and $T^2$ spatial topology a close relation to our discussion  of axion/dilaton gravity can be found. The main difference is that in 2+1 gravity the description can be exact essentially because gravitons are absent. Also instead of matter fields one uses the anisotropy parameters of the spatial metric.

The action for $2+1$ dimensional gravity is given by:
 
\[
S = \int {d^3 x\sqrt { - g} (M_{Pl} R - \frac{\lambda }{2})} 
\]
In $2+1$ dimensions one can work in special gauge where the metric components only depend on time. In this gauge the metric is written as:

\[
d\ell ^2  =  - Ndt^2  + V(t)\left( {\begin{array}{*{20}c}
   {dx} & {dy}  \\
\end{array}} \right)\frac{1}{{m_2 (t)}}\left( {\begin{array}{*{20}c}
   {m_1^2 (t) + m_2^2 (t)} & {m_1 (t)}  \\
   {m_1 (t)} & 1  \\
\end{array}} \right)\left( {\begin{array}{*{20}c}
   {dx}  \\
   {dy}  \\
\end{array}} \right)
\]
Here $m_1,m_2$ are anisotropy parameters but they essentially play the role of axion and dilaton zero modes in section 3 so we use the same notation.
Defining $\tilde N = NV$ and rescaling the Planck mass the action takes the form:
\[
S = \frac{1}{2}\int {dt\tilde N( - M_{Pl} \frac{{\dot V^2 }}{{\tilde N^2 }}}  + M_{Pl} V^2 \frac{{\dot m_1^2  + \dot m_2^2 }}{{m_2^2 \tilde N^2 }} - \lambda )
\]
In this expression we see that the the action corresponds to the action of a particle in background metric in $(V,m_1,m_2)$ space given by:

\[
\delta s^2  =  - M_{Pl} dV^2  + M_{Pl} V^2 \frac{1}{{m_2^2 }}(dm_1^2  + dm_2^2 )
\]
The constraint derived from the action is:

\[
- \pi _V^2  + \frac{1}{{V^2 }}m_2^2 (\pi_1^2  + \pi_2^2 ) + \lambda M_{Pl}=0
\]
Then upon quantization the Wheeler-Dewitt equation for $2+1$ gravity becomes an ordinary differential equation.

\[
( \frac{1}{{M_P }}\frac{\partial^2}{\partial V^2}  -  \frac{m_2^2}{{M_{Pl}V^2 }}(\frac{\partial^2}{\partial m_1^2}+  \frac{\partial^2}{\partial m_1^2})  + \lambda)\Phi = 0
\]
Following references \cite{McGuigan:1989tn}\cite{Waldron:2004gg} we make  the change of variables to:

\[
\begin{array}{l}
 X_1  = \frac{{\sqrt V }}{{2\sqrt {M_{Pl} } }}(\frac{{m_1^2  + m_2^2 }}{{m_2 }} - \frac{1}{{m_2 }}) \\ 
 X_2  = \sqrt V \frac{{m_1 }}{{\sqrt {M_{Pl} } m_2 }} \\ 
 X_0  = \frac{{\sqrt V }}{{2\sqrt {M_{Pl} } }}(\frac{{m_1^2  + m_2^2 }}{{m_2 }} + \frac{1}{{m_2 }}) \\ 
 \end{array}
\]
Then the metric in $(V,m_1,m_2)$ space becomes:

\[
\delta s^2  =  - dX_0^2  + dX_1^2  + dX_2^2 
\]
and the WDW equation then takes the simple form:
\[
( - P_0^2  + P_1^2  + P_2^2  + \lambda )\Phi  = 0
\]
In these variables it is very straightforward to see that $\lambda$ plays the role of $Mass^2$ in the Klein-Gordon type equation.

The transformation properties of the wave functions under the $SL(2,Z)$ symmetry of the anisotropy parameters are somewhat easier to describe in terms of the $(V,m_1,m_2)$variables so again by using separation of variables the wave functions are:
 
\[
\Phi(V,m_1,m_2) =\Phi_s(V)f_{\nu}(m_1,m_2) \sim H_{i\sqrt{\nu^2+\frac{1}{4}}}^{(2)} (\sqrt{\lambda}V\sqrt{M_{Pl}})f_{\nu}(m_1,m_2)
\]
At large $V$ these wave functions behave as:
\[
\Phi_s(V)\sim \exp(-iV\sqrt{M_{Pl}}\sqrt{\lambda})
\]
The formulation of $2+1$ gravity recounted above is usually used to describe a single universe. If one wants to describe multiple universes one can form a direct sum of $T^2$ spaces. However we are interested in direct sums which describe universes with different cosmological constants. In higher dimensions the vacuum energy on different spatial topologies can in principle be calculated and one obtains different Casimir energies which, if they come from internal dimensions, effectively act as different cosmological constants. In  $2+1$ gravity one seeks to mimic this but one doesn't want to introduce a fluctuating field to calculate the vacuum energy as this would destroy the solvability of the the theory. Thus one simply introduces two different cosmological constants directly into the theory. Doing this in the context of this paper we write a coupled set of WDW equations as:

\[
\begin{array}{l}
 (\frac{1}{{M_{Pl} }}\frac{\partial }{{\partial V}}\frac{\partial }{{\partial V}} + \frac{s(1-s)}{{M_{Pl}V^2 }} + \lambda _1  )\Phi _1  + \sqrt {\lambda _1 \lambda _2 } \Phi _2  = 0 \\ 
 (-\frac{1}{{M_{Pl}}}\frac{\partial }{{\partial V}}\frac{\partial }{{\partial V}} + \frac{s(1-s)}{{V^2 }} + \lambda _1  + \lambda _2  )\Phi _2  + \sqrt {\lambda _1 \lambda _2 } \Phi _1  = 0 \\ 
 \end{array}
\]
Again solving these coupled equations as before: 

\[
\begin{array}{l}
 \Phi _1 \sim H_{i\sqrt{\nu^2+ \frac{1}{4}}}^{(2)} (\sqrt {\lambda _ -  M_{Pl}} V) + \sqrt {\frac{{\lambda _1 }}{{\lambda _2 }}} H_{i\sqrt{\nu^2+ \frac{1}{4}}}^{(2)} (\sqrt {\lambda _ +  M_{Pl}} V) \\ 
 \Phi _2 \sim  - \sqrt {\frac{{\lambda _1 }}{{\lambda _2 }}} H_{i\sqrt{\nu^2+ \frac{1}{4}}}^{(2)}(\sqrt {\lambda _ -  M_{Pl}} V) + H_{i\sqrt{\nu^2+ \frac{1}{4}}}^{(2)}(\sqrt {\lambda _ +  M_{Pl}} V) \\ 
 \end{array}
\]
The large $V$ behavior of the solutions which again go as :
 
\[
\begin{array}{l}
 \Phi _1 \sim \exp( - i\sqrt{M_{Pl}} \sqrt {\lambda _ -  } V) +  \ldots  \\ 
 \Phi _2 \sim \exp( - i\sqrt{M_{Pl}} \sqrt {\lambda _ +  } V) +  \ldots  \\ 
 \end{array}
\]
with $\lambda_-$ and $\lambda_+$ given by the seesaw relations (2.5). Although the formulas are essentially the same as section 3 the main difference is that in $2+1$ dimensions there are no gravitons to modify the minisuperspace description of the wave functions and potentially effect the seesaw relationship.

\subsubsection{OSV Wave functions and Supergravity}

Another context in which minisuperspace can be exact are certain compactified string models considered in \cite{Ooguri:2005vr}.  This is related to the approach of Witten \cite{Witten:1981nf}\cite{Witten:1982df} who realized that when looking at topological quantities in supersymmetric models that involve low lying states one could obtain results much more easily using a simpler $0+1$ or $1+1$ field theory instead of the full theory. When generalized to gravitational interactions as in \cite{Ooguri:2005vr} the $0+1$ becomes is an exact description of the wave function by minisuperspace. For a $1+1$ reduction we would have an exact description in terms of midisuperspace \cite{McGuigan:1990nd}\cite{Macias:2005vu}.  Note also that the Wheeler De Witt equation has no time and can be viewed as a sort of zero energy condition of the form $H\Psi = 0$ \cite{Pioline:2002qz}.

The exact wave function found in \cite{Ooguri:2005vr} has arguments given by $X^I$ and ${\bar X}^I$ which are three form fluxes of the internal 6d Calabi Yau manifold, $K(X,\bar X)$ is the exponentiated Kahler potential and the $W(X)$ is the superprepotential. 
The Wheeler-DeWitt equation was reduced to two separate BPS type equations using supersymmetry and $H = \bar{Q} Q$.  The BPS equations were given by:

\[
\begin{array}{l}
 Q\Phi = (\frac{\partial }{{\partial X^I }} - \bar \partial _I K + i\bar \partial _I \bar W)\Phi  = 0 \\ 
 \bar{Q}\Phi =(\frac{\partial }{{\partial \bar X^I }} - \partial _I K - i\partial _I W)\Phi  = 0 \\ 
 \end{array}
\]
with the exact solution:

\[
\Phi(X,\bar X) = \exp (K(X,\bar X) + i(W(X) - \bar W(\bar X)))
\]

In this context the cosmological constant seesaw it is not too difficult to form as a direct sum of two such spaces and again we write a coupled set of WDW equations of the form:

\begin{equation}
\begin{array}{l}
 H_1 \Phi _1  =  - \Lambda _{12} \Phi _2  \\ 
 H_2 \Phi _2  =  - \Lambda _{12} \Phi _1  \\ 
  \\ 
 \end{array}
\end{equation}
In equation (5.1)  the cosmological constant of each universe is built in to the definition of $H_1$ and $H_2$ and so is not separated out from the superpotential and that is why only the mixing term $\Lambda_{12} $  is displayed. 
The main difficulty is that any seesaw relationship we have found is of the form $\lambda_1^2/\lambda_2$ so that at least one of the universes used in the seesaw should have positive cosmological constant if one want to have a cosmologically interesting prediction. However the discussion of \cite{Ooguri:2005vr} was strictly for $\lambda$ negative.  Nevertheless there may be a generalization to positive $\lambda$.  Such scenarios typically involve uplifting the superpotential by the addition of a constant term \cite{Buchbinder:2004im}\cite{Abe:2005pi}. If this can be developed in the context of OSV wave functions one should be able to build a cosmological seesaw using the exact wave functions.

Finally in four dimensional supergravity one can also write the constraints in terms of two  equations \cite{D'Eath:1984sh}:
 
\[
\begin{array}{l}
 S\Psi  = 0 \\ 
 \bar S\Psi  = 0 \\ 
 \end{array}
\]
with

\[
\begin{array}{l}
 S^{A'}  = \varepsilon ^{ijk} e_{AA'i} (x)D_j \chi _k^A (x) - \frac{1}{2}\chi _i^A \frac{\delta }{{\delta e_i^{AA'} (x)}} \\ 
 \bar S^A  = D_j \frac{\delta }{{\delta \chi _i^A (x)}} + \frac{1}{2}\frac{\delta }{{\delta e_i^{AA'} (x)}}D_{ij}^{A'B} \frac{\delta }{{\delta \chi _j^B (x)}} \\ 
 \end{array}
\]
and where $e$ and $\chi$ are spatial component of the vierbein and gravititino field respectively. The differential operator  $ D_{ij}^{A'B}$ can be found in \cite{D'Eath:1984sh}.

Now we couple a pair such supergravity constraint equations together in order to induce mixing between universes. Describing universes by wave functions $\Psi_1$ and $\Psi_2$ but  with different values of $\lambda$ we write:
\[
S\Psi_1  = (\sqrt {\lambda_1 } \varepsilon^{ijk} e_i^a e_j^b \gamma_a \gamma_b \chi_k )\Psi_2 
\]
\[
S\Psi_2  = (\sqrt {\lambda_1 } \varepsilon^{ijk} e_i^a e_j^b \gamma_a \gamma_b \chi_k )\Psi_1  + (\sqrt {\lambda_2 } \varepsilon^{ijk} e_i^a e_j^b \gamma_a \gamma_b \chi_k )\Psi_2 
\]
In this form we see that it again is the $M$ matrix:

\[
M = \left( {\begin{array}{*{20}c}
   0 & {\sqrt {\lambda _1 } }  \\
   {\sqrt {\lambda _1 } } & {\sqrt {\lambda _2 } }  \\
\end{array}} \right)
\]
which appears. So the generalization to supergravity is close to usual seesaw mechanism except the spinor components of the wave function of the universe play the role of the neutrino field. Finally a  mass for the gravitino also modifies this supergravity constraint equations equation in a similar way to $\sqrt{\lambda}$ \cite{Deser:1977ur}. However scenarios of light gravitino and heavy gravitinos are still possible \cite{Ibe:2005xc},
\cite{Endo:2005uy}. Thus there is as yet no clear relation  between the $m_{3/2}$ and the measurement of positive $\lambda$.

\section{Conclusion}

In this paper we have introduced a realization of a cosmological constant seesaw mechanism in quantum cosmology. The main observation we used was that the cosmological constant plays the role of a $Mass^2$ type term in the Wheeler-Dewitt equation. As each universe only has one cosmological constant a way to introduce a cosmological constant into a $2\times 2$ matrix is through mixing with other universes. One concrete way of introducing this mixing is by coupling two Wheeler-DeWitt equations together. If one of the universes before mixing has vacuum energy at the Planck scale  and one has  vacuum energy before mixing at the supersymmetry breaking scale we obtain the cosmological constant seesaw relationship. The mechanism differs from Bousso and Polchinki in that the anthropic principle is not needed. Intriguing derivations of exact wave functions of the universe in string theory may allow for a more precise realization  of the cosmological seesaw mechanism in the future.


\begin{thebibliography}{100}


\bibitem{Motl}
http://motls.blogspot.com/2005/12/cosmological-constant-seesaw.html

\bibitem{Carroll}
http://cosmicvariance.com/2005/12/05/duff-on-susskind/

\bibitem{McGuigan:1989yb}
  M.~McGuigan,
  ``Universe Decay And Changing The Cosmological Constant,''
  Phys.\ Rev.\ D {\bf 41}, 418 (1990).

\bibitem{Russo:2004am}
  J.~G.~Russo and P.~K.~Townsend,
  ``Cosmology as relativistic particle mechanics: From big crunch to big
  bang,''
  Class.\ Quant.\ Grav.\  {\bf 22}, 737 (2005)
  [arXiv:hep-th/0408220].

\bibitem{Townsend:2004zp}
  P.~K.~Townsend and M.~N.~R.~Wohlfarth,
  ``Cosmology as geodesic motion,''
  Class.\ Quant.\ Grav.\  {\bf 21}, 5375 (2004)
  [arXiv:hep-th/0404241].


\bibitem{Fischler:1989ka}
  W.~Fischler, I.~R.~Klebanov, J.~Polchinski and L.~Susskind,
  ``Quantum Mechanics Of The Googolplexus,''
  Nucl.\ Phys.\ B {\bf 327}, 157 (1989).
\bibitem{Carlip:1991ij}
  S.~Carlip,
  ``(2+1)-dimensional Chern-Simons gravity as a Dirac square root,''
  Phys.\ Rev.\ D {\bf 45}, 3584 (1992)
  [Erratum-ibid.\ D {\bf 47}, 1729 (1993)]
  [arXiv:hep-th/9109006].

\bibitem{Ooguri:2005vr}
  H.~Ooguri, C.~Vafa and E.~P.~Verlinde,
  ``Hartle-Hawking wave-function for flux compactifications,''
  Lett.\ Math.\ Phys.\  {\bf 74}, 311 (2005)
  [arXiv:hep-th/0502211].

\bibitem{Craps:2005wd}
  B.~Craps, S.~Sethi and E.~P.~Verlinde,
  ``A matrix big bang,''
  JHEP {\bf 0510}, 005 (2005)
  [arXiv:hep-th/0506180].

\bibitem{Maharana:2004qs}
  J.~Maharana,
  ``Symmetries of axion-dilaton string cosmology,''
  Int.\ J.\ Mod.\ Phys.\ A {\bf 20}, 1441 (2005)
  [arXiv:hep-th/0405039].


\bibitem{McGuigan:1990pi}
  M.~McGuigan,
  ``Constraints For Toroidal Cosmology,''
  Phys.\ Rev.\ D {\bf 41}, 3090 (1990).

\bibitem{Smith:1991up}
  E.~Smith and J.~Polchinski,
  ``Duality survives time dependence,''
  Phys.\ Lett.\ B {\bf 263}, 59 (1991).



\bibitem{Gasperini:1991ak}
  M.~Gasperini and G.~Veneziano,
  ``O(d,d) covariant string cosmology,''
  Phys.\ Lett.\ B {\bf 277}, 256 (1992)
  [arXiv:hep-th/9112044].
\bibitem{Bousso:2000xa}
  R.~Bousso and J.~Polchinski,
  ``Quantization of four-form fluxes and dynamical neutralization of the
  cosmological constant,''
  JHEP {\bf 0006}, 006 (2000)
  [arXiv:hep-th/0004134].

\bibitem{Horava:1996ma}
  P.~Horava and E.~Witten,
  ``Eleven-Dimensional Supergravity on a Manifold with Boundary,''
  Nucl.\ Phys.\ B {\bf 475}, 94 (1996)
  [arXiv:hep-th/9603142].
\bibitem{Randall:1999ee}
  L.~Randall and R.~Sundrum,
  ``A large mass hierarchy from a small extra dimension,''
  Phys.\ Rev.\ Lett.\  {\bf 83}, 3370 (1999)
  [arXiv:hep-ph/9905221].

\bibitem{Duncan:1990fr}
  M.~J.~Duncan and L.~G.~Jensen,
  ``A New Mechanism For Neutralizing The Cosmological Constant,''
  Nucl.\ Phys.\ B {\bf 361}, 695 (1991).

\bibitem{Narain:1986gd}
  K.~S.~Narain and M.~H.~Sarmadi,
  ``Removing Tachyons By Compactification,''
  Phys.\ Lett.\ B {\bf 184}, 165 (1987).
\bibitem{Wilczek:2005aj}
  F.~Wilczek,
  ``Enlightenment, knowledge, ignorance, temptation,''
  arXiv:hep-ph/0512187.

\bibitem{McG} M.~McGuigan,``Noncritical M-theory: Three dimensions,'' \\
arXiv:hep-th/0408041.

\bibitem{Hor}
  P.~Horava and C.~A.~Keeler,
  ``Noncritical M-theory in 2+1 dimensions as a nonrelativistic Fermi liquid,''
  arXiv:hep-th/0508024.

\bibitem{Par}
  J.~H.~Park,
  ``Noncritical osp(1$|$2,R) M-theory matrix model with an arbitrary time
  dependent cosmological constant,''
  arXiv:hep-th/0510070.


\bibitem{McGuigan:1991qp}
  M.~D.~McGuigan, C.~R.~Nappi and S.~A.~Yost,
  ``Charged black holes in two-dimensional string theory,''
  Nucl.\ Phys.\ B {\bf 375}, 421 (1992)
  [arXiv:hep-th/9111038].
\bibitem{Davis:2005qi}
  J.~L.~Davis,
  ``The moduli space and phase structure of heterotic strings in two
  dimensions,''
  arXiv:hep-th/0511298.
\bibitem{Davis:2005qe}
  J.~L.~Davis, F.~Larsen and N.~Seiberg,
  ``Heterotic strings in two dimensions and new stringy phase transitions,''
  JHEP {\bf 0508}, 035 (2005)
  [arXiv:hep-th/0505081].
\bibitem{Lukas:1997fg}
  A.~Lukas, B.~A.~Ovrut and D.~Waldram,
  ``On the four-dimensional effective action of strongly coupled heterotic
  string theory,''
  Nucl.\ Phys.\ B {\bf 532}, 43 (1998)
  [arXiv:hep-th/9710208].

\bibitem{Carlip:1994tt}
  S.~Carlip and R.~Cosgrove,
  ``Topology change in (2+1)-dimensional gravity,''
  J.\ Math.\ Phys.\  {\bf 35}, 5477 (1994)
  [arXiv:gr-qc/9406006].
\bibitem{Martinec:1984fs}
  E.~J.~Martinec,
  ``Soluble Systems In Quantum Gravity,''
  Phys.\ Rev.\ D {\bf 30}, 1198 (1984).

\bibitem{McGuigan:1989tn}
  M.~D.~McGuigan,
  ``Third Quantization And Quantum Cosmology,''
UMI-90-16127 (1989)
\bibitem{Waldron:2004gg}
  A.~Waldron,
  ``Milne and torus universes meet,''
  arXiv:hep-th/0408088.


\bibitem{Witten:1981nf}
  E.~Witten,
  ``Dynamical Breaking Of Supersymmetry,''
  Nucl.\ Phys.\ B {\bf 188}, 513 (1981).

\bibitem{Witten:1982df}
  E.~Witten,
  ``Constraints On Supersymmetry Breaking,''
  Nucl.\ Phys.\ B {\bf 202}, 253 (1982).


\bibitem{McGuigan:1990nd}
  M.~McGuigan,
  ``The Gowdy Cosmology And Two-Dimensional Gravity,''
  Phys.\ Rev.\ D {\bf 43}, 1199 (1991).

\bibitem{Macias:2005vu}
  A.~Macias, H.~Quevedo and A.~Sanchez,
  ``Gowdy T**3 cosmological models in N = 1 supergravity,''
  arXiv:gr-qc/0505013.


\bibitem{Pioline:2002qz}
  B.~Pioline and A.~Waldron,
  ``Quantum cosmology and conformal invariance,''
  Phys.\ Rev.\ Lett.\  {\bf 90}, 031302 (2003)
  [arXiv:hep-th/0209044].



\bibitem{Buchbinder:2004im}
  E.~I.~Buchbinder,
  ``Raising anti de Sitter vacua to de Sitter vacua in heterotic M-theory,''
  Phys.\ Rev.\ D {\bf 70}, 066008 (2004)
  [arXiv:hep-th/0406101].

\bibitem{Abe:2005pi}
  H.~Abe, T.~Higaki and T.~Kobayashi,
  ``Moduli-mixing racetrack model,''
  arXiv:hep-th/0512232.

\bibitem{D'Eath:1984sh}
  P.~D.~D'Eath,
  ``The Canonical Quantization Of Supergravity,''
  Phys.\ Rev.\ D {\bf 29}, 2199 (1984)
  [Erratum-ibid.\ D {\bf 32}, 1593 (1985)].


\bibitem{Deser:1977ur}
  S.~Deser, J.~H.~Kay and K.~S.~Stelle,
  ``Hamiltonian Formulation Of Supergravity,''
  Phys.\ Rev.\ D {\bf 16}, 2448 (1977).



\bibitem{Ibe:2005xc}
  M.~Ibe, K.~Tobe and T.~Yanagida,
  ``A gauge-mediation model with a light gravitino of mass O(10)-eV and the
  messenger dark matter,''
  Phys.\ Lett.\ B {\bf 615}, 120 (2005)
  [arXiv:hep-ph/0503098].


\bibitem{Endo:2005uy}
  M.~Endo, M.~Yamaguchi and K.~Yoshioka,
  ``A bottom-up approach to moduli dynamics in heavy gravitino scenario:
  Superpotential, soft terms and sparticle mass spectrum,''
  Phys.\ Rev.\ D {\bf 72}, 015004 (2005)
  [arXiv:hep-ph/0504036].




 \end{thebibliography}
\end{document}